\documentclass[twoside,twocolumn,9pt]{article}
\usepackage{extsizes}
\usepackage[super,sort&compress,comma]{natbib} 
\usepackage[version=3]{mhchem}
\usepackage[left=1.5cm, right=1.5cm, top=1.785cm, bottom=2.0cm]{geometry}
\usepackage{balance}
\usepackage{times,mathptmx}
\usepackage{sectsty}
\usepackage{graphicx} 
\usepackage{lastpage}
\usepackage[format=plain,justification=justified,singlelinecheck=false,font={stretch=1.125,small,sf},labelfont=bf,labelsep=space]{caption}
\usepackage{float}
\usepackage{fancyhdr}
\usepackage{fnpos}
\usepackage[english]{babel}
\addto{\captionsenglish}{%

}
\usepackage{array}
\usepackage{droidsans}
\usepackage{charter}
\usepackage[T1]{fontenc}
\usepackage[usenames,dvipsnames]{xcolor}
\usepackage{setspace}
\usepackage[compact]{titlesec} 
\usepackage{hyperref}

\definecolor{cream}{RGB}{222,217,201}

\begin{document}

\pagestyle{fancy}
\thispagestyle{plain}
\fancypagestyle{plain}{
}

\makeFNbottom
\makeatletter
\renewcommand\LARGE{\@setfontsize\LARGE{15pt}{17}}
\renewcommand\Large{\@setfontsize\Large{12pt}{14}}
\renewcommand\large{\@setfontsize\large{10pt}{12}}
\renewcommand\footnotesize{\@setfontsize\footnotesize{7pt}{10}}
\makeatother

\renewcommand{\thefootnote}{\fnsymbol{footnote}}
\renewcommand\footnoterule{\vspace*{1pt}%
\color{cream}\hrule width 3.5in height 0.4pt \color{black}\vspace*{5pt}} 
\setcounter{secnumdepth}{5}

\makeatletter 
\renewcommand\@biblabel[1]{#1}            
\renewcommand\@makefntext[1]%
{\noindent\makebox[0pt][r]{\@thefnmark\,}#1}
\makeatother 
\renewcommand{\figurename}{\small{Fig.}~}
\sectionfont{\sffamily\Large}
\subsectionfont{\normalsize}
\subsubsectionfont{\bf}
\setstretch{1.125} 
\setlength{\skip\footins}{0.8cm}
\setlength{\footnotesep}{0.25cm}
\setlength{\jot}{10pt}
\titlespacing*{\section}{0pt}{4pt}{4pt}
\titlespacing*{\subsection}{0pt}{15pt}{1pt}


\makeatletter 
\newlength{\figrulesep} 
\setlength{\figrulesep}{0.5\textfloatsep} 

\newcommand{\topfigrule}{\vspace*{-1pt}%
\noindent{\color{cream}\rule[-\figrulesep]{\columnwidth}{1.5pt}} }

\newcommand{\botfigrule}{\vspace*{-2pt}%
\noindent{\color{cream}\rule[\figrulesep]{\columnwidth}{1.5pt}} }

\newcommand{\dblfigrule}{\vspace*{-1pt}%
\noindent{\color{cream}\rule[-\figrulesep]{\textwidth}{1.5pt}} }

\makeatother

\twocolumn[
  \begin{@twocolumnfalse}
\vspace{3cm}
\sffamily
\begin{tabular}{m{0.5cm} p{16.5cm} }

& \noindent\LARGE{\textbf{Multi-scale theoretical approach to X-ray absorption spectra in disordered systems: an application to the study of Zn(II) in water}} \\
\vspace{0.3cm} & \vspace{0.3cm} \\

& \noindent\large{Francesco Stellato,\textit{$^{a,g}$} Matteo Calandra,\textit{$^{b}$} Francesco D'Acapito,\textit{$^{c}$} Emiliano De Santis,\textit{$^{d}$} Giovanni La Penna,\textit{$^{a,e}$} Giancarlo Rossi,\textit{$^{a,d,f}$} and Silvia Morante,\textit{$^{a,d}$}} \\


& \noindent\normalsize{{

We develop a multi-scale theoretical approach aimed at calculating from first principles X-ray absorption spectra of liquid solutions and disordered systems. We test the method by considering the paradigmatic case of Zn(II) in water which, besides being relevant in itself, is also of interest for biology. With the help of classical molecular dynamics simulations we start by producing bunches of configurations 
differing for the Zn(II)-water coordination mode. Different coordination modes 
are obtained by making use of the so-called dummy atoms method. 
From the collected molecular dynamics trajectories, snapshots of a more manageable subsystem encompassing the metal site and two solvation layers are cut out. Density functional theory is used to optimize and relax these reduced system configurations employing a uniform dielectric to mimic the surrounding bulk liquid water. On the resulting structures, fully quantum mechanical X-ray absorption spectra calculations are performed by including core-hole effects and core-level shifts. The proposed approach does not rely on any guessing or fitting of the force field or of the atomic positions of the system. The comparison of the theoretically computed spectrum with  the experimental Zn K-edge XANES data unambiguously demonstrates that among the different {\it a priori} possible geometries, Zn(II) in water lives in an octahedral coordination mode. 
}}


\end{tabular}

 \end{@twocolumnfalse} \vspace{0.6cm}
]

\renewcommand*\rmdefault{bch}\normalfont\upshape
\rmfamily
\section*{}
\vspace{-1cm}


\footnotetext{\textit{$^{a}$~INFN, Sezione di Roma 2, Via della Ricerca Scientifica, I-00133 Roma, Italy}}
\footnotetext{\textit{$^{b}$~Sorbonne Universit\'e, CNRS, Institut des Nanosciences de Paris, UMR7588, F-75252, Paris, France}}
\footnotetext{\textit{$^{c}$~CNR-IOM-OGG c/o European Synchrotron Radiation Facility, 71 Avenue des Martyrs, F-38043 Grenoble, France}}
\footnotetext{\textit{$^{d}$~Dipartimento di Fisica, Universit\`a di  Roma
  ``Tor Vergata'', Via della Ricerca Scientifica, I-00133 Roma, Italy}}
\footnotetext{\textit{$^{e}$~CNR - Institute for Chemistry of Organometallic Compounds, Sesto Fiorentino, 50019, Italy}}
\footnotetext{\textit{$^{f}$~Centro Fermi - Museo Storico della Fisica e Centro Studi e Ricerche ``Enrico Fermi'', Rome, Italy}}
\footnotetext{\textit{$^{g}$~Corresponding author, E-mail: stellatof@roma2.infn.it}}




\section{Introduction}
\label{sec:INTRO}

Zinc is the second (after iron) most abundant transition metal in living organisms. It is known to carry out a number of important different tasks in complex with proteins. A detailed knowledge of zinc ion chemistry is essential for understanding its role in biology and for designing either complexes that deliver zinc to proteins or chelating agents that instead remove zinc from proteins~\cite{krkezel2016biological}.

A necessary premise to all these investigations is of course the identification of the Zn(II) coordination mode in water. The structure of this system is in principle quite simple, and yet it is still under debate in the literature.

The most commonly observed Zn(II)-water coordination mode is the hexa-coordinated octahedral one~\cite{mhin1992zn,mink2003infrared,rudolph1999zinc}. However, penta- and tetra-coordinated structures have been also proposed~\cite{arumuganathan2008two,bock1995hydration,krkezel2016biological,Dudev00} and are observed when Zn is in complex with proteins or in the presence of solvents different from pure water~\cite{auld2001zinc,jacquamet1998x,d2002total,giannozzi2012zn}.

The relevance of non-octahedral coordination geometries is still a matter of debate (see Ref.~\citenum{krkezel2016biological} for a survey). Moreover, the presence of a small fraction of \ce{Zn(OH)^+} aqua-ions is possible, even at physiological conditions,  because at pH=7.4 one expects that in about 1\% of the cases the Zn(II) ion is coordinated to an OH$^-$ ion. Depending on the nature of the anion X in the \ce{ZnX2} salt that is dissolved in water and of its concentration, the presence of \ce{ZnX_n|^{(2-n)+}}, $n=1,2,3$, species has been proposed since a long time~\cite{Maeda96}. We see from this discussion that, even the coordination mode of a simple system like Zn(II) in water solution can be affected by the existence of several other species, all probed by spectroscopy. Thus a self-consistent computational tool able to directly model the spectrum and to include all these effects is necessary to be able to reliably settle the question. 

Various experimental~\cite{mink2003infrared,migliorati2012influence} and computational~\cite{sanchez1996examining,mohammed2005quantum} techniques have been used to study the coordination of Zn(II) ions in water. Among the experimental techniques, X-ray absorption spectroscopy (XAS) is the ideal tool for probing the local environment around a selected atom in a disordered system. XAS works, in fact, for systems in any state of aggregation and can therefore be used for structural investigations of metal ions in solution. Another remarkable feature of this technique is that the choice of the XAS absorption edge allows for chemical and orbital selectivity. This ``tunability'' has led to the development of X-ray microscopy as a powerful analytical technique, with the possibility of detailed space resolved measurements (imaging) even for trace concentrations, as it is often the case in biology~\cite{Collingwood2017}.

A XAS spectrum is commonly divided into two regions according to the kinetic energy of the extracted electron: the so called XANES (X-ray Absorption Near Edge Structure) region that conventionally extends from a few eV before the edge energy (in the case of the K-edge the latter is the ionization energy of the 1s electron of the absorbing atom) and few tenths of eV above it, and the EXAFS (Extended X-ray Absorption Fine Structure) region that starts after the XANES portion of the spectrum and extends 500-1000 eV after the edge. 

The XANES region, in which the extracted electron undergoes mainly multiple scattering events with the surrounding atoms, in principle contains detailed information about the local atomic structure around the absorbing atom~\cite{filipponi1995x,benfatto1986multiple,rehr2005progress, morante2014metals}. However, in order to extract geometrical information from the XANES, a quite accurate guess of the actual geometrical arrangement of the atoms surrounding the absorber must be available. This is necessary because, given the low kinetic energy of the extracted electron, its behaviour strongly depends on the details of lattice and core-hole potentials in the vicinity of the absorbing atom. 

Various approaches aimed at extracting valuable structural information from the XANES spectrum have been developed and are routinely used ~\cite{benfatto2003mxan,bunuau2009self,rehr2010parameter}. Some of them (e.g.\ the one implemented in the popular MXAN code~\cite{benfatto2003mxan}), make use of the so-called ``muffin-tin'' approximation, which consists in assuming a spherical scattering potential centered on each atom and a constant value in the interstitial region among atoms. A different class of approaches is based on first principles electronic structure computations either via the calculation of the real-space Green function~\cite{rehr2010parameter} or exploiting plane waves basis sets and pseudopotentials~\cite{taillefumier2002x, gougoussis2009intrinsic,gougoussis2009first,Bunau2013}. The approach we shall present and discuss in the work belongs to this last class of methods.

On the computational side, classical and quantum-mechanical methods have been employed either alone~\cite{mohammed2005quantum,riahi2013qm,sanchez1996examining,merkling2001combination,merkling2002exploring,minicozzi2008role} or in combination with experimental information to determine the structures of ions in solution. Synergic methods of this kind rely on the idea of producing via molecular dynamics (MD) simulations a pool of individual configurations of the system representative of static disorder effects. The average over the XANES spectra computed for each one of the collected configurations is in the end compared to experimental data.

This kind of strategy was also employed in Ref.~\citenum{migliorati2012influence}, 
where Zn(II) structures in water are constructed from classical MD simulations guided by a force field in turn fitted on an underlying density-functional theory (DFT) calculation. The MXAN software, employing the muffin-tin potential, is finally used to compute the theoretical spectrum.

In this paper we present a parameter-free first principles calculation of the XANES spectrum of Zn(II) in water similar to the one that has been already successfully developed in Ref.~\citenum{la2015first} for Cu(II) in water. 
In the present case we improved our XANES calculation strategy by taking care of very relevant aspects, such as the core-level shift calculation (see section~\ref{sec:CXS} for the details), that were not dealt with in the previous work~\cite{la2015first}.
We proceed by first generating by classical MD equilibrated configurations of Zn(II) in a large box of water molecules.
From the large system snapshots, configurations of a smaller subsystem including the metal site and two solvation layers are cut out. The latter are then relaxed within the DFT formalism suitable for large (100-1000) atomic assemblies. The relaxation step is intended to drive to acceptable values the forces resulting from the extraction of a sub-set of atoms from the larger system and the quantum mechanical inclusion of valence electrons (that we represent in terms of plane waves) and core pseudopotentials.

Our calculations were performed in a plane wave formalism. The main advantages of this method with respect to local basis sets are twofold. 
First, structural optimization is more robust in a plane waves formalism as there are no Pulay forces and it 
is not necessary to displace the localized sets. Thus, even for very large atomic displacement and optimized
structures very different from the initial guess, the optimization step remains very reliable. This is crucial in liquids, where the atoms are 
substantially free to move and large statistic is required to compare calculations with experiments~\cite{liang2011ab}.
Second, atomic orbitals are more inaccurate as the energy of the XAS spectrum increases (far-edge region) and the final
state is closer to a scattering state (i.e.\ a plane wave).
Finally, from the computational point of view, our approach relies on the Lanczos method and continuous fractions which only require the knowledge of the electronic charge density and avoid the cumbersome explicit calculation of excited states. This represents a substantial computational gain as neither the band energies nor the wavefunctions of the empty electronic
state are needed. This aspect is important when dealing with relatively large cells (about 100 atoms), which, as observed in the literature~\cite{mo2000ab}, are required to properly calculate XANES spectra.
The XANES spectrum is therefore here computed by performing a fully self-consistent DFT calculation based on plane waves and the use of pseudopotentials~\cite{taillefumier2002x,gougoussis2009intrinsic,gougoussis2009first,Bunau2013}. The comparison of the computed spectrum with the experimental Zn K-edge XANES data unambiguously shows that among the different {\it a priori} possible geometries, Zn(II) in water lives in an octahedral coordination mode

Methodologically the computational strategy we present in this work is an important step in the direction of an unbiased, quantum-mechanical framework aiming at a parameter-free  calculation of the XANES spectra of disordered systems. The simplicity and the success of the strategy we have illustrated and tested in this work make us to believe that it can be exported to more complicated situations like the one we encounter in disordered systems and in systems of biological relevance.

 \section{Materials and Methods}
 \label{sec:MandM}

\subsection{Experimental data}

The experimental spectrum which we will compare our theoretical calculation with was measured at the GILDA beamline at ESRF~\cite{d1998gilda}. The monochromator was equipped with two Si(311) crystals and was run in the sagittally focusing configuration. Collimation and harmonic rejection was achieved via two Pd coated mirrors working at 3 mrads (cutoff energy = 22 keV). XAS data were collected in fluorescence mode using a 7 elements high-purity-germanium detector. Zn solution was obtained by dissolving 2 mM of ZnCl$_2$ salt in deionized water. Data were collected at room temperature.

\subsection{Building empirical Zn(II) coordination geometries}

The first step of our computational strategy consists in the construction of representatives of the {\it a priori} possible coordination geometries of the metal ion in water. 
In the case of Zn(II) the relevant metal coordination modes we need to focus on are the octahedral, tetrahedral and square-planar configurations. It should be noticed that the somewhat unusual square-planar coordination was considered to model the approximate Zn penta-coordination arising when one water molecule approaches the plane formed by the 4 nearest Zn ligand water molecules from an axial direction (see the discussion in Sect.~\ref{sec:Res}). 

A practical way to build the desired geometrical water arrangements around the metal site is to make recourse to the ``dummy atoms'' method~\cite{pang2000successful}.
The method consists in placing at fixed positions positive charges (referred to as dummy atoms) around the metal ion center taken as neutral. The location and the charges of the dummy atoms are chosen so as to match the selected geometry, octahedral, tetrahedral and square-planar, respectively, with the Zn charge distributed over them~\footnote{We note incidentally that the charge smearing provided by dummy atoms helps avoiding possible spurious effects due to a too high charge density concentration at the metal site.}. By running a sufficiently long MD simulations (1~ns turns out to be a sufficiently long time) dummy charges have the effect of  ``gently'' pushing the water molecules surrounding the metal site to take their positions in the desired geometry. 

In the case of the octahedral coordination 6 dummy atoms each of charge $e/3$ ($e$ is the electron charge) are located at the vertices of an octahedron with its center on the Zn site. In the case of the square-planar coordination, we have 4 dummy atoms of charge $e/3$ each sitting at the vertices of a square with its center on the Zn site. The tetrahedral coordination is construct in a way similar to the square-planar one, but with the dummy atoms lying at the vertices of a tetrahedron with Zn at the center~\footnote{It should be observed that in the cases of the square-planar and tetrahedral geometries the total charge of the dummy atoms is smaller than the nominal Zn ion charge, $Q_{Zn(II)}=2|e|$. This is not a problem as we are here only preparing initial configurations of Zn in water with the metal ion in the desired coordination mode that will be then relaxed by DFT methods (see next section) with each atom having its correct electron number.}.

The total charge of the cell was kept equal to zero by adding an appropriate number of Cl$^-$ counterions~\cite{Aqvist90}. In all the geometries we have considered the dummy atoms are located at a distance of 0.9~\AA\ from the Zn site.

The interactions between Zn and TIP3P water molecules are described by a Lennard--Jones  potential. The Lennard-Jones parameters of the Zn-O interaction are taken to be $\sigma=2.49$~\AA\ and $\epsilon=2.77$ cal/mol. The simulation box was obtained starting from a cubic box (with side $L=18.774$~\AA) of TIP3P water molecules~\cite{jorgensen1983comparison} endowed with periodic boundary conditions and equilibrated by Monte Carlo methods at normal conditions ($T=300$~K and $p=1$~bar), where 3 water molecules were replaced by a Zn atom and two chloride anions (for charge neutralization). The resulting system is composed by one Zn atom, two chloride anions and 213 water molecules. Owing to periodic boundary conditions electrostatic interactions were computed employing the smooth particle-mesh Ewald (SPME) method~\cite{essmann1995smooth}. We took a cut-off of 9~\AA, a mesh grid with 0.1~\AA\ spacing and a PME direct space energy tolerance of $10^{-4}$ kcal/mol. Classical MD simulations in the $N,V,T$ {\it ensemble} were run with the help of the NAMD code~\cite{phillips2005scalable} with a time step of 2~fs. The temperature was kept fixed at $T= 300$~K by using the stochastic Berendsen thermostat~\cite{berendsen1984molecular}. Bond length constraints for bonds involving H atoms was enforced with the help of the SHAKE algorithm~\cite{Ryckaert77}.

\subsection{Density functional theory calculations}

As we discussed in the previous section, in order to sample the system configuration space, we have proceeded by selecting for each Zn coordination geometry (octahedral, tetrahedral and square-planar) an adequately large number of configurations along the collected MD trajectories. These will be used for the successive calculation of the XANES spectrum. 

We have found that 20 configurations per each one of the three Zn coordination modes we are considering are enough. As we shall see, in fact, with this number we are able to capture the system thermal and structural fluctuations as they are probed by the XANES spectrum. Adding further configurations on the other hand improves neither the quality of the theoretical spectrum nor its statistical significance. In practice we have selected one configuration every 50~ps along the 1~ns long MD trajectories. 

For the actual first principles XANES calculation we are aiming at, the systems we have constructed are too big. Taking advantage of the fact that for the determination of the XANES spectrum only features within the second Zn solvation shell are relevant, we have cut out from the whole system a sphere around the metal ion including up to the second solvation shell. An analysis of the Zn-O radial distribution function on the configurations collected along the simulated MD trajectories shows that on average 29 water molecules are contained in this sphere. Noting that none of the selected configurations displays a chloride anion within 6~\AA\ from Zn, the final systems we will be considering for the successive analysis are just made by the Zn ion plus the 29 water molecules closest to Zn.

These reduced system are inserted in a cubic super-cell box with a side of 2.2~nm with the ``empty'' space filled with a uniform dielectric mimicking the bulk liquid water at standard conditions~\cite{andreussi2012revised}. At this point all the configurations need to be relaxed to eliminate unphysical strains and contacts. This is done with the help of the Broyden--Fletcher--Goldfarb--Shanno quasi-Newton algorithm~\cite{Fletcher00} by minimizing the system potential energy. 

The latter is quantum-mechanically evaluated in the DFT formalism using the {\sc QuantumESPRESSO} (v.\ 5.0.2) suite of programs~\cite{giannozzi2009quantum,Giannozzi2017}. We use Vanderbilt ultra-soft pseudopotentials~\cite{Vanderbilt90} with inclusion of semicore states for Zn and the PBE exchange-correlation functional~\cite{Perdew96}. Electronic wave functions were expanded in plane waves up to an energy cutoff of 50~Ry, while a 500~Ry energy cutoff was used for the expansion of the charge density. The $\Gamma$-point sampling in reciprocal space was adopted in all the electronic structure calculations.

It should be noted that the relaxation step which all the selected configuration are subjected to is not intended for bringing the system to its absolute (global) minimum which strictly speaking would represent the uninteresting $T=0$ system configuration~\footnote{If any, one would like to minimize the free energy of the system and not its potential energy.}. The decrease in potential energy is mostly due to relaxation of bond distances towards their equilibrium values with the non-bonding interactions assisting these local changes. Within the number of relaxation steps that we have employed we witness a change in the Zn coordination mode only in the case of the metastable square-planar coordination that in a number of instances turn into a square base pyramidal or an octahedral coordination. The octahedral and tetrahedral Zn coordination modes look instead quite stable local minima of the system configuration space. 

In Table~\ref{tab:Tab1} we compare the structural parameters before (i.e.\  in the configurations extracted from the MD simulation) and after the DFT relaxation step for the three bunches of configurations of the reduced system we have considered, 
The Table is divided in two parts. The left (columns 1 to 5) and right (columns 6 to 10) part refer to the structures before and after the DFT relaxation, respectively. In the parts of the Table, in the first column we indicate the Zn coordination mode at the beginning and at the end of the relaxation procedure with in parentheses the number of configurations having the same geometry we have considered. As we remarked above, only in the case of the initial square-planar geometry the Zn coordination mode evolves into different structures. In 12 cases Zn ends up in a square base pyramidal coordination, in 5 cases in an octahedral one and in the remaining 3 cases the Zn coordination remains unmodified. In the second and third column of each half of the Table we report the radius of the first and second Zn solvation shell averaged over the set of configurations with the same geometry. The error is taken as half the dispersion of the values computed over each set of configurations. It should be noted that the dispersion of the values of the mean radius of the first solvation shell ($\Delta r^{(1)}$) is definitely smaller than for the dispersion of the mean radius of the second solvation shell ($\Delta r^{(2)}$), both before and after the relaxation step. This should be interpreted as a sign of the variability of the second solvation shell even among configurations having the same geometry. One also notices that the errors obtained after the DFT relaxation are always larger than those coming directly out from MD simulations indicating a larger geometrical variability of configurations after DFT relaxation.

In the fourth column of each part we show the value of the Bond Valence Sum (BVS)~\cite{brown1981bond} averaged over the configurations of each type of Zn coordination. BVS is an empirical parameter that, when charges and structural parameters of nearby atoms are correctly assigned, should come out to be of the order of magnitude of the nominal ion oxydation state, which is 2 in the case at hand. It is worth pointing out that, while the BVS values obtained for the MD structures are always higher than the expected value of 2, those calculated for the DFT relaxed  structures are in agreement with the expected one. This confirms that the DFT relaxation is a relevant step to obtain reliable structural parameters.

Finally, in the last column we give the values of the Zn charge computed according to the Natural Orbital Population (NOP) analysis~\cite{Reed85}. The calculation was performed with the help of the Gaussian code~\cite{Gaussian16,G16manual} using the hybrid M06-2X exchange functional~\cite{Zhao08} for Zn. The DFT set-up emploies a localized basis-set of type 6-31+G(d) for N, O, C and H atoms, and LANL2DZ for Zn, the latter including a pseudopotential approximation for Zn core electrons. The NOP analysis is little sensitive to the choice of the basis-set. Solvent effects are modelled by means of the implicit polarizable continuum model of Ref.~\citenum{Marenich09}. Fixed atomic coordinates corresponding to the MD and DFT relaxed structures which Table~\ref{tab:Tab1} refers to are used in the NOP electronic calculations. 

The Zn charges obtained from the present analysis are consistent with previous calculations of Zn(II) in water in octahedral coordination performed in~\cite{Pokherel2016}. As observed, the amount of positive charge transfer from the metal ion to ligands decreases when water molecules are added in the second shell at fixed coordination geometry. 
Indeed, in Ref.~\citenum{Pokherel2016} it was found that the charge on Zn increases from 1.4 to 1.7, upon adding up to three water molecules to the second shell. In our calculation, where the whole second shell is included, the positive charge is, instead, more delocalized as a consequence of using the PBE exchange functional. Since the decrease of charge transfer occurs mainly because the number of ligands decreases, as expected, the increase of the Zn charge in our case is smaller than in the case considered in Ref.~\citenum{Pokherel2016}. This is in accordance with the fact that in the latter case only a few water molecules (up to three) were added to the second shell, while in our case a fully occupied second shell of water molecules is always part of the system.

\begin{table*}[ht] 
\begin{center}
\small
\begin{tabular}{|c|c|c|c|c||c|c|c|c|c|}
\hline
\multicolumn{5}{|c||}{MD structures} & \multicolumn{5}{c|}{DFT relaxed structures}\\
\hline
Geometry & $r^{(1)} \pm \Delta r^{(1)}$ & $r^{(2)} \pm \Delta r^{(2)}$ & BVS & Zn charge & Geometry & $r^{(1)} \pm \Delta r^{(1)}$ & $r^{(2)} \pm \Delta r^{(2)}$ & BVS & Zn charge \\
\hline
{\it oct} (20) & 1.93 $\pm$ 0.05 & 4.2 $\pm$ 0.2 & 3.2  $\pm$ 0.2& 1.51$\pm$0.01 & {\it oct} (20) & 2.12 $\pm$ 0.08 & 4.2 $\pm$ 0.3 & 2.0  $\pm$ 0.1& 1.51$\pm$0.01 \\
\hline
{\it tth} (20) & 1.86 $\pm$ 0.06 & 4.1 $\pm$ 0.2  & 2.6  $\pm$ 0.2& 1.62$\pm$0.01 & {\it tth} (20) &  1.96 $\pm$ 0.06 & 4.1 $\pm$ 0.3  & 2.0  $\pm$ 0.1 & 1.62$\pm$0.01\\
\hline
& & & & & {\it sqp} (3)& 2.01 $\pm$ 0.04  & 4.2 $\pm$ 0.3  & 1.8 $\pm$ 0.1 & 1.68$\pm$0.02 \\
{\it sqp} (20)& 1.95 $\pm$ 0.07  & 4.2 $\pm$ 0.2  & 2.4 $\pm$ 0.1 &1.58$\pm$0.05 & {\it sqp} (12) &  2.06 $\pm$ 0.09 &  4.2 $\pm$ 0.3 & 1.9  $\pm$ 0.1 & 1.58$\pm$0.01 \\
& & & & &  {\it oct} (5)& 2.13 $\pm$ 0.09  &  4.2 $\pm$ 0.3 & 1.9  $\pm$ 0.1 & 1.52$\pm$0.01  \\
\hline
\end{tabular}
\caption{\small{The ``evolution'' of the structural parameters from the configurations extracted along the MD simulation (left half of the table) to the configurations after the DFT relaxation step (right half of the table). The quantities $r^{(1)}$ and $r^{(2)}$ are the mean distances between Zn and water oxygen atoms in the first and second solvation sphere, respectively, averaged over the set of configurations whose number is reported in parentheses. Errors on distances are taken as half the dispersion of the mean values for each geometry. In the last two columns of each half we report the values of the Bond Valence Sum (BVS)~\cite{brown1981bond} and of the Zn charge, calculated using the NOP procedure~\cite{Reed85}, averaged over each set of  configurations. Errors are statistical. We use the following abbreviations: octahedral = {\it oct}; tetrahedral = {\it tth}; square planar = {\it sqp}; square base pyramidal = {\it sqbp}. 
\label{tab:Tab1}}
}
\end{center}
\end{table*} 

\begin{figure}[h]
\centering
\includegraphics[width=0.3\textwidth]{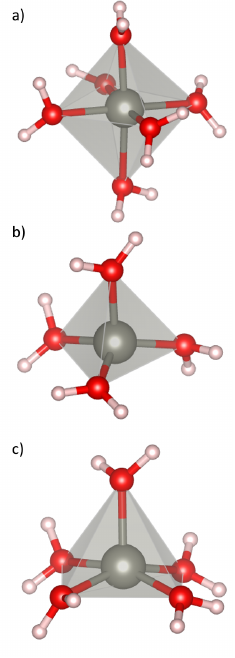}
 \caption{A {\it Ball\&Stick} representation of Zn environment in water as it results after the DFT relaxation step for the octahedral (panel a)) tetrahedral (panel b)) and penta-coordinated (panel c)) mode.}
\label{fig:fig1}
\end{figure}

In Fig.~\ref{fig:fig1} we show examples of the geometry of the Zn(II)-water environment in the case of the octahedral (panel a)), tetrahedral (panel b)) and penta-coordinated mode (panel c)). 

\subsection{Computation of the XANES spectrum}
\label{sec:CXS}

We have computed the XANES spectrum of the 20 relaxed octahedral and tetrahedral configurations listed in Table~\ref{tab:Tab1} using the {\sc XSPECTRA}~\cite{taillefumier2002x,gougoussis2009intrinsic,gougoussis2009first,Bunau2013} package of the {\sc QuantumESPRESSO} suite.  As for the square-planar coordination, the calculation of the XANES spectrum was carried out only for the (more numerous) 12 final square based pyramidal (penta-coordinated) configurations resulting from the relaxation of the square planar geometry. 

The theoretical XANES spectra that will be finally compared with the experimental data are the averages taken over the set of configurations belonging to each coordination geometry (octahedral, tetrahedral and square based pyramidal).

The XANES calculation was performed in the dipole approximation. Core-hole effects were taken into account by generating a Zn pseudopotential with a hole in the 1s state. The spectra are convoluted~\cite{krause1979natural,bunuau2013projector} with a Lorentzian having an arctangent-like, energy-dependent width $\Gamma$. The minimum of $\Gamma$, attained at low energy and up to the Fermi energy, is taken to be 1.7~eV, while its maximum, reached at high (infinite) energy is 4~eV.  The inflection point is located at 30~eV. 

\begin{figure}[h]
\centering
\includegraphics[height=6cm]{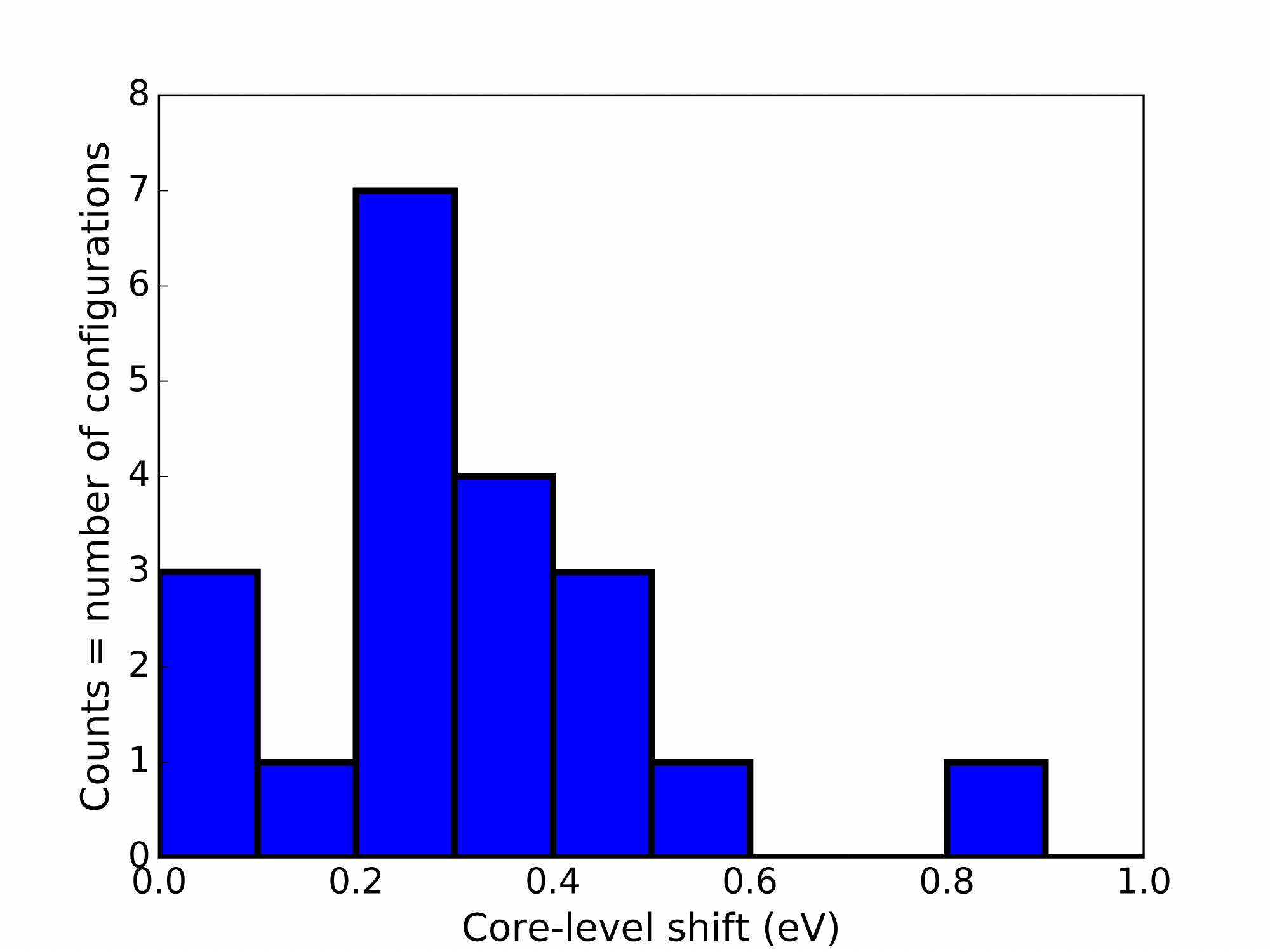}
 \caption{Histogram of the core-level shifts magnitude for the 20 octahedral configurations of Table~\ref{tab:Tab1} after the relaxation step.}
\label{fig:fig2}
\end{figure}

In the presence, as it is the case here, of inequivalent absorbing sites in the unit cell, the value of the energy, $E_i$, of the initial state depends on the geometry around the absorbing site due to the core-level shift. Thus, in order to have a sensible comparison among the theoretically computed XANES spectra for each geometry so as to be able to correctly match the energies in performing the average, the spectra of each set need to be ``re-aligned'' by explicitly calculating the core level shift as explained in Refs.~\citenum{gao2008theory,nemausat2015phonon, jiang2013experimental, mizoguchi2009first, lelong2014detecting}. More precisely, we first align the energies of the lowest unoccupied level of each configuration, $\epsilon_{LUB}$, and then we displace the whole spectrum by the difference $E_{TOT}^{*}-E_{TOT}$, where $E_{TOT}^{*}$ and $E_{TOT}$ are the total energies of the given configuration in the presence and in the absence of a core-hole, respectively~\cite{nemausat2015phonon}~\footnote{The core level shift  is calculated according to the formula (see eq.~(22) of Ref.~\citenum{nemausat2015phonon})
\begin{eqnarray}
E \rightarrow E - \epsilon_{LUB} + (E_{TOT}^{*} - E_{TOT})\, .\nonumber
\end{eqnarray}
In this shift the energy of the lowest unoccupied electronic band, $\epsilon_{LUB}$, is subtracted and the total energy difference between the system with one 1s core hole plus one electron in the first available electronic state ($E_{TOT}^{*}$) and that of the ground state ($E_{TOT}$) is added.}. 
This correction is often claimed to be negligible and ignored. An explicit calculation for the three sets of configurations we considered shows that this is not the case here. For example, in the case of octahedral configurations, the core-level shift can be as large as about 1~eV, as shown in the histogram of Fig.~\ref{fig:fig2}. 

 \section{Results}
 \label{sec:Res}

As we said, in Table~\ref{tab:Tab1} we summarize the geometrical modifications occurring at the end of the DFT relaxation step to the three (octahedral, tetrahedral and square-planar) sets of 20 configurations extracted from the 1~ns classical MD simulations described in Sect.~\ref{sec:MandM}. 

The first very clear (and not unexpected) result is that the square-planar configurational appears to be quite unstable. In fact, only 3 (out of 20) configurations maintain the square-planar geometry after the DFT relaxation step. Note also that in these three cases the BVS value is significantly lower than its nominal value. Among the other square-planar initial configurations, in 12 cases a fifth water molecule is attracted within the Zn coordination sphere and in 5 cases a further extra water molecule approaches the ion. Our simulation data show that the instability of the square-planar geometry is not to be ascribed to a too small number of ligands, but rather to the requirement of a spherically symmetric ligand arrangement, as forced by the spherical symmetry of the d$^{10}$ electronic
structure of \ce{Zn^{2+}}. We see, in fact, that starting from a tetrahedral configuration, which features the same number of ligands as the square-planar configuration, not only this coordination mode is maintained, but the corresponding BVS value is like what we expect it to be.

However, as already mentioned in the  previous section, we stress that none of the initially tetrahedral coordinations changes its coordination number. This is due to the interactions between the first and the second solvation layers (see also below) that, despite the low energy involved in hydrogen bonds, do not allow the displacement of water molecules from the second to the first Zn coordination sphere. This lack of mobility is a usual effect in energy relaxation in systems containing many water molecules extracted from a larger liquid sample. 

The situation for the octahedral geometry is similar to what we see in the tetrahedral case. The octahedral geometry looks very stable and the only visible effect of the DFT relaxation is a slight increase of the ligands--Zn mean distance. 

It is worth noticing that including the second solvation shell of Zn in the DFT model allows to properly take into account the interactions between the Zn-bound water molecules and the environment of the complex in solution. For the average distance between O belonging to water molecules in the Zn first solvation shell and O belonging to water molecules in the Zn second solvation shell, $\langle r^{(1,2)} \rangle$, we find in the octahedral geometry case of $\langle r^{(1,2)} \rangle|_{Zn\,in\,water} = 2.7 \pm 0.3$~\AA. 
This value is smaller than the value obtained by averaging the O-O distances contributing to the first peak ($r \le 3.5$~\AA) in the radial distribution function of the TIP3P water molecules, for which one finds $\langle r^{(1,2)} \rangle|_{pure\,water} = 3.0\pm0.3$~\AA.
As already observed in other calculations ~\cite{migliorati2012influence,sanchez1996examining,Smirnov2013},
the hydrogen bond between Zn-bound water molecules and water molecules in the second solvation shell is stronger than the hydrogen bonds of liquid water. This difference confirms that, as expected, the Zn-bound water molecules are interacting via activated hydrogen bonds with nearby water molecules. These interactions affect the electron ground state around the Zn center and, thanks to the quantum mechanical DFT treatment used in our approach, they are fully included in the simulation of the XANES spectra.

The {\sc XSPECTRA} code is employed to compute the XANES spectrum of each one of the configurations belonging to the three geometries listed in the second column of Table~\ref{tab:Tab1} (20 configurations for the octahedral and tetrahedral geometry, and 12 for the square base pyramidal one (penta-coordinated)).

\begin{figure*}
\centering
\includegraphics[height=5.0cm]{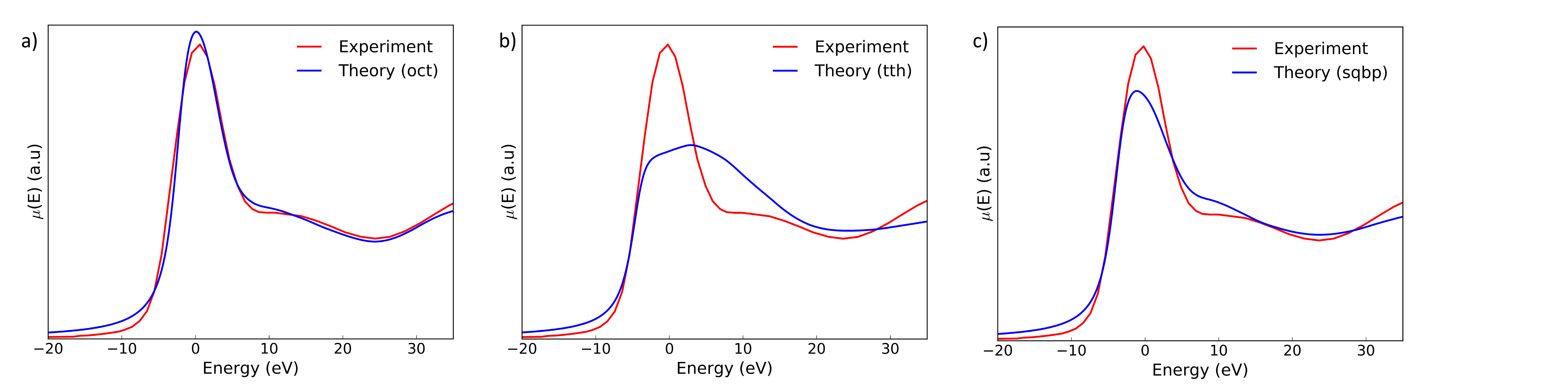}
 \caption{Octahedral (panel a)), tetrahedral (panel b)) and penta-coordinated (panel c)) simulated spectra (blue curves) compared with the experimental data acquired at the GILDA beamline at ESRF~\cite{d1998gilda} (red curve).} 
\label{fig:fig3}
\end{figure*}

As described in Sect.~\ref{sec:CXS}, each spectrum computed in this way is appropriately shifted and convoluted with a Lorentzian. Then, an average of the shifted and convoluted spectra is performed within each class of coordination geometry. The three average spectra are finally compared to experimental data taken at the GILDA beamline at ESRF~\cite{d1998gilda}. The comparison is shown in Fig.~\ref{fig:fig3} where in red we have drawn the experimental curve and in blue the theoretical ones.

We clearly see that the octahedral geometry is in very good agreement with the experimental XANES spectrum. On the contrary, the simulated spectra from the tetrahedral and the penta-coordinated geometries strongly deviate from the experimental one. This conclusion can be made more quantitative by computing and comparing of the $R$-factors of the three sets of data~\footnote{The $R$-factor is an estimate of the similarity between two sets of data. It is defined by the formula
\begin{eqnarray}
R= \frac{\sum_{i=1}^N \vert \mu ^{ex} (E_i) - \mu ^{th} (E_i) \vert}{\sum_{i=1}^N \vert \mu ^{ex} (E_i)\vert} \nonumber
\end{eqnarray}
where $\mu^{ex} (E_i)$ and $\mu^{th} (E_i)$ are the two sets of data, typically $\mu^{ex} (E_i)$ and $\mu^{th} (E_i)$ are the experimental and the simulated/theoretical  data, respectively.}. They are collected in Table~\ref{tab:tab3}.

\begin{table}
\begin{center}
\begin{tabular}{|l|c|}
\hline
Coordination mode & $R$-factor \\
\hline
\hline
octahedral  &  0.06 \\
\hline
tetrahedral & 0.19 \\ 
\hline
penta-coordinated  & 0.09 \\
\hline
\end{tabular}
\caption{\small{$R$-factors comparing experimental data and simulated spectra for  the octahedral, tetrahedral and square base pyramidal (penta-coordinated) geometry, respectively.}}
\label{tab:tab3}
\end{center}
\end{table}
One gets $R= 0.06$ for the octahedral coordination mode, $R = 0.19$ for the  tetrahedral one and  $R = 0.09$ for the penta-coordinated mode. These numbers confirm the qualitative observation already drawn by looking at Fig.~\ref{fig:fig3} that the coordination mode of Zn in water is octahedral. This conclusion is certainly not unexpected. But the key point of the strategy we have presented is that this result has been obtained in a fully first principles approach with no free parameters and no fitting. 


\section{Conclusions}

Extending the general strategy we have developed in the case of Cu in water~\cite{la2015first}, we have been able to accurately reproduce the XANES spectrum of Zn(II) in water, proving that among different {\it a priori} plausible geometries, Zn in water lives in an octahedral coordination.

The main virtue of the approach we present in this paper is that the calculation of the XANES spectrum is performed from first principles in a completely parameter-free way. 

Rather good fits of the XANES spectrum of Zn in water are already available in the literature~\cite{migliorati2012influence,d2002combined}. However, in those works, some kind of fit against a variable number of nonstructural parameters was performed. A parameter-free approach, like the one advocated in this paper, is instead aimed at calculating XANES spectra of complex systems of interest, not only in biology, but also in material science, without {\it ad hoc} assumptions or fitting ans{\"a}tze.

\section*{Conflicts of interest}
There are no conflicts to declare.

\section*{Acknowledgements} The calculations have been performed under the agreement between INFN and the National Supercomputing Consortium of Italy CINECA. The authors thank Y.\ Joly for useful discussions. This work was partly supported by INAIL grant BRiC 2016 ID17/2016.


\balance


\bibliography{Stellato_etal_1409}
\bibliographystyle{rsc} 

\end{document}